# Enabling the Ambient Pressure Growth of $ScB_2$ Crystals for AlGaN Power Electronics

Satya K. Kushwaha,[1,2,3,*] Ahamed Raihan,[4] Astrid D. Kengne,[4] Daniel Joel Harrison,[4] R. Shipra,[4] Han Xie,[1,2] Evan N. Crites,[4] Allana G. Iwanicki,[2] Luke J. Meiler,[2] Sharad Mahatara,[5] Maxime A. Siegler,[2] Renae N. Gannon,[5] Steven R. Spurgeon,[5,6,7] Rajeswari Kolagani,[8] Amitayush Jhu Thakur,[9] Stephan Lany,[5] Patrick E. Hopkins,[10] Jessica L. McChesney,[9] Nancy Haegel,[5] Tyrel M. McQueen,[1,2,11] Michael G. Spencer,[4] MVS Chandrashekhar[4]

*[1]William H. Miller III Department of Physics & Astronomy, Johns Hopkins University, Baltimore, MD 21218, USA*

*[2]Department of Chemistry, Johns Hopkins University, Baltimore, MD 21218, USA*

*[3]Department of Physics, University of North Texas, Denton, TX 76203, USA*

*[4]Department of Electronics and Computer Engineering, Morgan State University, Baltimore, MD 21251, USA*

*[5]National Laboratories of the Rockies, Golden, CO 80204, USA*

*[6]Metallurgical and Materials Engineering, Colorado School of Mines, Golden, CO 80401, USA*

*[7]Renewable and Sustainable Energy Institute (RASEI), University of Colorado Boulder, Boulder, CO 80309, USA*

*[8]Department of Physics, Astronomy and Geosciences, Towson University, Towson, MD 21252, USA*

*[9]Argonne National Laboratory, Lemont, IL 60439, USA*

*[10]Mechanical and Aerospace Engineering, The University of Virginia, Charlottesville, VA 22904, USA*

*[11]Department of Materials Science and Engineering, Johns Hopkins University, Baltimore, MD 21218, USA*

## Abstract

Here we report the growth of single crystalline $ScB_2$, an ultrahigh-temperature-ceramic, at ambient pressure in a laser-heated Optical Floating Zone via the travelling solvent method. Crystals have been grown from both Sc-rich (55-65 at% Sc) and B-rich self-flux (80-83 at% B) at growth rate in the range of 0.2 – 2 mm/hr. The structure of grown crystals is in good agreement with an $AlB_2$ type layered hexagonal phase, space group of P6/mmm, and lattice constants are $a$ = 3.1423 (2) Å (resp. 3.1502(3) Å) and $c$ = 3.5084(3) Å (resp. 3.5041(3) Å) for the crystals grown in Sc (resp. B-rich) conditions. Crystals natively grow along the in-plane [100] direction. Electron backscattered diffraction shows that Sc-flux results in boules with multiple domains with Sc inclusions, with the crystal-domains highly aligned. In contrast, B-flux boules are single domain after an initial nucleation region. Rocking curves measurements of B-flux crystals for the two reflection orders for ($h$000) and (000$l$) planes show single peaks, establishing that the crystals are free from grain boundaries, but with an asymmetry in the scattered intensity along the tails is suggestive of the presence of point defects. Surface X-ray photoemission spectroscopy measurements show that the electronic environment in B-flux crystals is much superior to that of Sc-flux ones and produced highly resolved binding energy peaks for B 1$s$ and Sc 2$p$. Work function measurements for the (11-20) plane is ~5 eV – supporting the highly electrically conductive nature of $ScB_2$. Our work demonstrates the viable growth of $ScB_2$, that it is an ideal lattice matched-substrate candidate and can be explored for the development of Al-rich AlGaN for power microelectronics, and the ambient pressure growth route providing the opportunity for scalable substrate manufacturing of $ScB_2$.

[*] To Whom Correspondence Should Be Addressed: Satya.Kushwaha@unt.edu

## Introduction:

Advancements in technology demand the discovery and synthesis of appropriate materials that enable the manufacturing of devices by exploiting the full efficiency of materials and manipulating them in a controlled way. $Al_{1-y}Ga_yN$, or AlGaN for short, is an ultrawideband gap (UWBG) semiconductor enabling future technologies, including power microelectronics and quantum information.[1–3] However, there is no lattice matched substrate for $Al_{1-y}Ga_yN$ in the y~0.3 - 0.5 range that is optimal for high power electronics. Therefore, its heteroepitaxy on available substrates grows with inevitable lattice defects including stacking faults and dislocations.[4–6] These extended lattice defects, combined with thermal and electrical resistance limitations of existing substrates, severely limit high voltage and high current device operation and restrict from tapping the full potential of AlGaN.[7–9] It is thus extremely important to explore new substrate modalities lattice matched to AlGaN with high electrical and thermal conductivity. To our knowledge only a few reports are available.[10–12] Some ultrahigh temperature ceramics (UHTC), specifically metal-diborides ($MB_2$; $M$ = Sc, Zr, Hf) and metal-carbides ($MC$, $M_2C$; $M$ = Ta, Nb), do lattice-match AlGaN, and their high electrical and thermal conductivities make them ideal substrates for fabricating highly efficient vertical power devices.[13–15] The (0001) surfaces of $ZrB_2$ crystals have been extensively explored as lattice matched substrate for GaN, and recently thin films growths of $ZrB_2$ have also been explored as virtual substrates and crystals were mainly grown by floating zone.[16–18] Recently, $Ta_{0.8}Hf_{0.2}C_{1-\delta}$ was proposed as a universal substrate for AlGaN.[19]

$ScB_2$, a metal-diboride UHTC,[20,21] is reported to crystallize in an AlB2-type hexagonal lattice symmetry with lattice constants $a$ = 3.148 Å, $c$ = 3.516 Å, and space-group $P6_3$/mmm (191).[22] The $a_{hex}$ of $ScB_2$ is lattice-matched with 50% Al AlGaN, in the middle of the range ideal for power electronics.[23,24] $ScB_2$ lattice matching provides an opportunity to deposit the würtzite phase AlGaN on either of Sc and B terminated surfaces with low misfit dislocation density, reduced mosaicity, and cracks, etc. Since $ScB_2$ is free from O therefore it also allows the growth of AlGaN at high temperature without introducing O defects. In $ScB_2$, the Sc hexagonal and B honeycomb layers are neutral with least dangling bonds – the B terminated layer directly provides an opportunity to form strong B–N bonds and stabilize the nucleation of N-polar, and Sc terminated surface may also structurally and chemically favor for Al-polar AlGaN by creating strong Sc-N bond with incoming N.[25,26] It is also a good electrical and thermal conductor therefore provides an opportunity to enhance the Baliga figure of merit of vertical devices by handling the operation of AlGaN at high voltages and high temperatures.[27] However, thus far there are no reports, other than our concurrent related work, on the fusing of $ScB_2$ with AlGaN for power microelectronics.[27]

$ScB_2$ is reported to melt congruently at ~2250 ºC, however, as the temperature approaches its melting temperature, a significant evaporation of Sc and $ScB_2$ creates difficulties in achieving a stoichiometric melt, and normally requires a high gas pressure in the growth chamber to suppress the Sc vaporization. So far, only one report is available on the growth of $ScB_2$ crystals from Sc-rich molten at high argon pressure of 1 MPa (10 bar).[24]

In this work, we present the growth and characterization of single-crystalline $ScB_2$ prepared from both the Sc- and B-rich self-flux molten zones at ambient pressure via a traveling solvent approach in a laser assisted floating zone furnace, described elsewhere.[28] Crystals grown from Sc-rich flux have significant Sc metal inclusions and multiple co-aligned domains, while B-rich flux results in high-quality single-domain crystals free from grain boundaries. Elemental composition and crystal structure have been analyzed in detail by energy dispersive spectroscopy and X-ray diffraction techniques. The crystalline quality is assessed by electron back-scattering diffraction and high-resolution x-ray diffraction. The surface electronic structure and chemical environment is studied by x-ray photoemission spectroscopy, and the room temperature electrical and thermal transport is measured, the essential properties for $ScB_2$ being a lattice-matched metallic substrate for AlGaN.

## Experimental

### Crystal Growth Process

Crystal growth was carried out in a diode-laser assisted floating zone (LDFZ) furnace (Crystal Systems, Inc., FD-FZ-5-200-VPO-PC). This system is equipped with five 200 W GaAs lasers (λ = 976 nm) with a tilt-option (0 – 35º upward in vertical-plane) for each laser, and all the laser beams are focused into a small region of ~4mm height and 8 mm width (Fig. 1 (a,b)). The laser heating provides a large temperature gradient at solid-liquid interface and allows controlled heating of smaller regions of the samples. This is extremely helpful for the crystal growth of materials which exhibit high evaporation at their melting points, such as $ScB_2$. Depending on the optical coupling of the laser light with the material, zone temperatures as high as 2600 ºC can be achieved. Commercial polycrystalline rods (5 mm diameter, 100 mm long) from Testbourne Ltd. were used as starting materials. The seed and feed rods were attached to the upper and lower shafts of the LDFZ furnace (Fig. 1(c)). Crystal growth was explored in two regions of the B – Sc phase diagram: (i) Sc-rich, and (ii) B-rich. For the growth from Sc-self flux, a Sc metal (99.99%, purchased from Kurt J. Lesker) piece of ~500 mg was placed on the top of lower (feed) rod to create traveling molten zone (represented by a green color region in Fig. 1(c)), with an estimated zone composition of 55-65 at% Sc. The growth chamber was flushed 3 times by successive evacuating and purging with Ar gas, and during growth process Ar flow rate was maintained at 1-1.5 L/min. The upper rod was used as seed rod for spontaneous nucleation and crystal growth; the end of the rod was made sanded in pencil shape to limit the nucleation in smaller region with fewer domain growth. Upon creating a stable zone, the seed/feed rods were counter rotated at 10/rpm, and the lasers were operated between 60-80 % of power throughout the experiment. The molten zone was created by melting Sc piece and slowly dissolving the $ScB_2$ rod into it. For Sc-rich growth, the molten zone (~55-65 at% Sc) was created by melting Sc piece on top of $ScB_2$ rod and slowly dissolving this $ScB_2$ rod into the molten Sc. A pellet of $B_{0.82}Sc_{18}$ composition (82 at% B) was placed on top of the $ScB_2$ feed rod to have a B-rich flux molten zone. Several crystals were successful pulled from Sc and B rich travelling molten zones at growth rates ranging from 1mm/hr to 1.5mm/hr and depending on flux compositions crystals are referred as SBS and SBB respectively for Sc and B-rich throughout the paper. The details on the characterization techniques and results on grown crystals are discussed in the following sections.

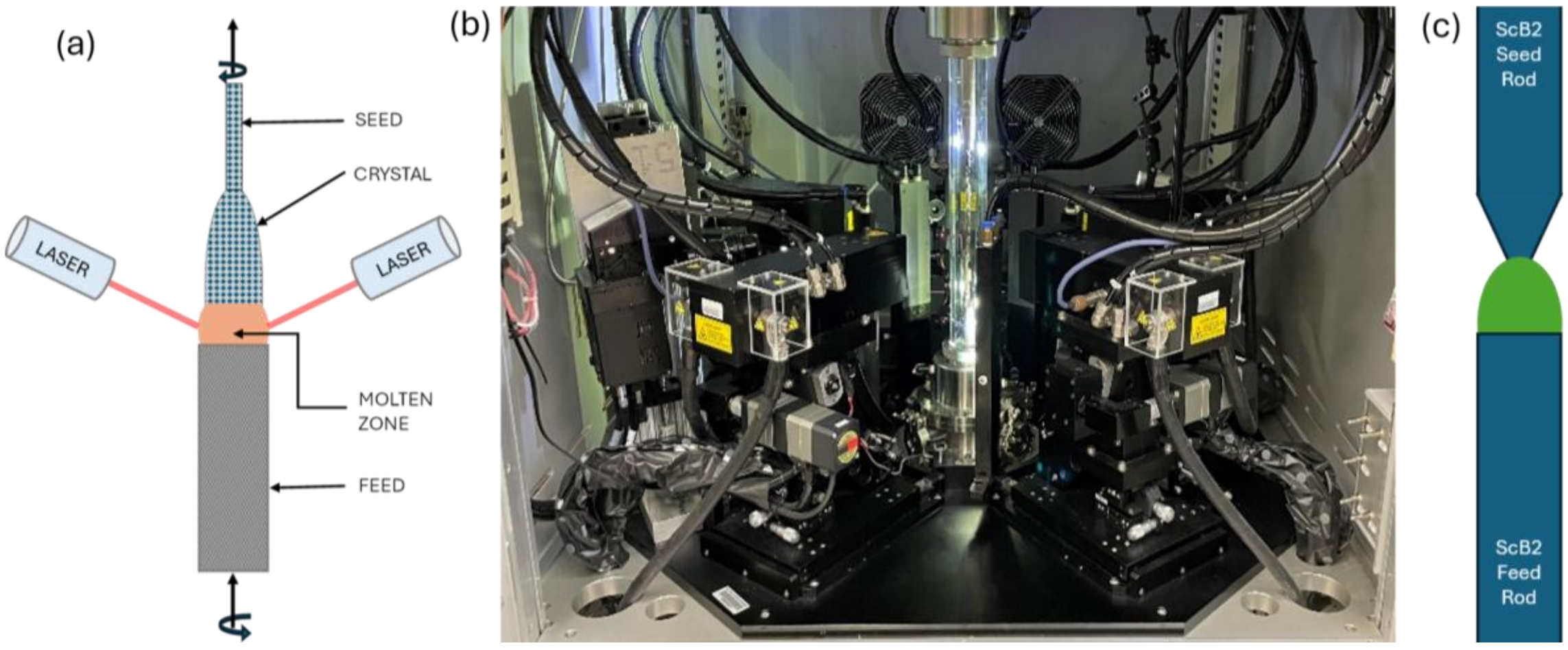

*Fig. 1. (a) The schematic depicting laser-pedestal growth geometry in laser heated floating zone growth with molten zone, feed & seed sample rods, laser heating source – the arrows indicate respective rotation and translation motions; (b) a photograph of the inner view of the laser-diode floating zone (situated in PARADIM Johns Hopkins University), clearly displaying 5 laser assembly and quartz-tube chamber for crystal growth; (c) schematic representation of the setup for the $ScB_2$ crystal growth – the molten zone flux (green) is held between two polycrystalline $scB_2$ rods, the upper rod acts as seed and lower one as feed rod. The tip of seed rod was tapered to reduce the number of spontaneous nucleation at the beginning of crystal growth.*

## Single crystal and powder X-ray diffraction

All reflection intensities were measured at $T$ = 213(2) K using a SuperNova diffractometer (equipped with Atlas detector) with Mo Kα radiation (λ = 0.71073 Å) under the program CrysAlisPro (Version CrysAlisPro 1.171.42.49, Rigaku OD, 2022). The same program was used to refine the cell dimensions and for data reduction. The structure was solved with the program SHELXT-2018/2 (Sheldrick, 2018) and was refined on F2 with SHELXL-2019/3 (Sheldrick, 2018). Numerical absorption correction based on gaussian integration over a multifaceted crystal model was applied using CrysAlisPro. The temperature of the data collection was controlled using the system Cryojet (manufactured by Oxford Instruments).

The full refinement data was collected on a small crystal of SBS with the size 0.16×0.09×0.05 (mm), and a total of 892 reflections were recorded. For the SBB crystal, minimal data sufficient for unit cell parameter determination was collected.

Room temperature powder XRD patterns were collected using a Bruker D8 Discover diffractometer that employs a Cu X-ray source with a wavelength, λ = 0.15406 nm, and a position sensitive lynx eye detector. Some samples were also measured on Bruker D8 Focus with a LynxEye detector, and a Ni filter to remove the Cu-K_β rays. The crystals were ground in the form of fine powder and spread on the sample holder in the form of a thin layer with the help of grease. Peak search and phase identification was performed with the help of Bruker's DIFFRAC.EVA software.

**X-ray Microcomputed Tomography**

A Bruker SkyScan 1172 system operating at 100 kV and 100 $\mu$A with downstream Al and Cu filters was used to collect a 360° dataset for boule SBS. Data was analyzed by Dragonfly software, Version 2021.1 for Windows (Object Research Systems Inc, Montreal, Canada, 2020; software available at http://www.theobjects.com/dragonfly).

**Laue diffraction**

Laue diffraction patterns on the grown crystals were obtained using a Multiwire Laboratories Ltd. MWL120 Real-Time Back-Reflection Laue Camera System controlled with a MWL 731 Six-Axis Motor Driver with a Spellman DF60N3 x-ray Generator providing an x-ray source. The generator was operated at 10 mA, between 10 and 15 kV, with 15 kV. NorthStar software was used to simulate and analyze the diffraction spots. The crystals were tested for the head-on growth as well as on the side-surfaces to identify the spontaneous growth direction of crystals.

**High resolution X-ray diffraction**

The structural quality of grown crystals was analyzed by high-resolution x-ray diffraction (HRXRD). The high-resolution $2\theta$–$\omega$ x-ray diffraction pattern and $\omega$ rocking curves on a ($h$000) and (000$l$) cut sample were recorded on the Rigaku SmartLab Studio high resolution diffractometer with a two-bounce monochromator (resolution of ~0.007°). The equipment broadening is 24 arcsec.

**Scanning electron microscopy**

Surfaces of SBS and SBB cleaved samples were prepared for electron microscopy by Argon Broad Ion Beam (BIB) surface polishing on a JEOL IB-19520CCP Cross Section Polisher at 6 kV for 15 minutes with constant 360-degree sample rotation at approximately a 12 degree incident angle to the surface. Plasma focused ion beam (PFIB) was then used to mill a rectangular border around large representative areas of the samples for correlative measurements. Elemental composition and distribution of these regions were analyzed by Energy Dispersive X-Ray Spectroscopy (EDS) using an EDAX Octane Elite Super EDS detector on a Thermo Fisher Scientific Helios 5 Laser Hydra. Electron backscatter diffraction (EBSD) of the Sc-rich sample was collected on an EDAX Clarity EBSD detector on a Thermo Fisher Scientific Helios 5 Laser Hydra with a 2 μm step-size. For the B-rich sample, an Oxford Symmetry EBSD detector on a Thermo Fisher Scientific Helios 5 CXe was used at a 3 μm step size.

### X-ray Photoemission Spectroscopy

X-ray photoemission spectroscopy (XPS) measurements were performed at beamline 29-ID of the Advanced Photon Source, Argonne National Laboratory. Synchrotron radiation with photon energy of 1000 eV was used as the excitation source. The emitted photoelectrons were analyzed using a Scienta R4000 hemispherical analyzer with a pass energy of 200 eV (with an experimental energy resolution of 200 meV, combined X-ray and analyzer). The Sc flux and $ScB_2$ crystallite measurements were taken on SBS samples, and the $ScB_2$ crystals measurements were taken on SBB samples. The sample temperature was maintained at < 80 K. The binding energies of core levels were calibrated with respect to the Fermi level. The SBS sample was annealed in vacuum at 600 C for 10 mins before measurement, and the SBB sample was cleaved in vacuum at 60 K.

## Results and Discussion

In agreement with prior reports, ambient pressure melt tests on stoichiometric $ScB_2$ did not result in a reasonable melt upon heating but instead significant evaporation. Clean melting was not observed even when thermally isolating the melt from the sample holder (Schematic in Fig. S1). EDS analysis indicates a significant loss of Sc (Fig. S2), and such non-stoichiometry would significantly affect physical and mechanical properties.[29]

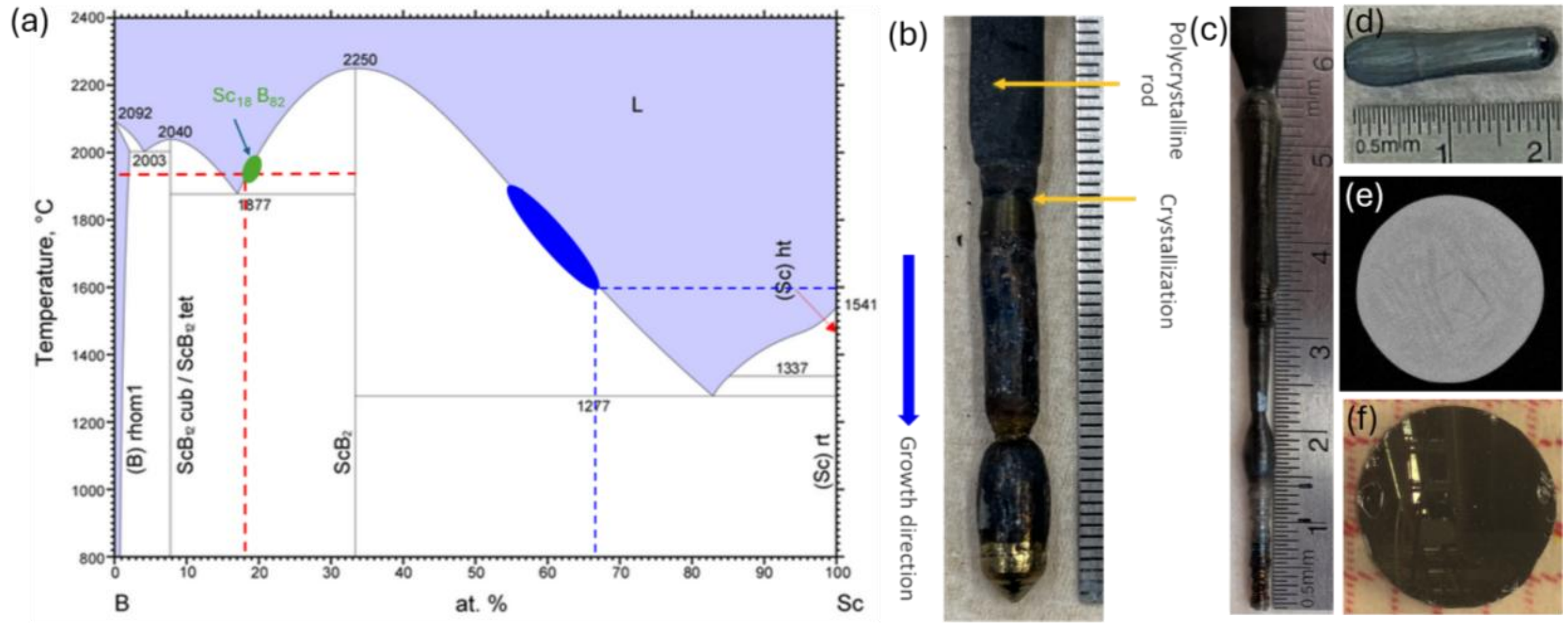

*Figure 2. (a) the B-Sc phase-diagram with labeled B-rich (green) and Sc-rich (blue) composition regions used for the self-flux travelling molten zones, the dotted lines are guide to the eye; (b) a representative crystal boule grown from Sc-rich molten zone, a blue arrow indicates the growth direction, and yellow arrows show the polycrystalline rod and spontaneous crystallization regions on the sample; (c, d) the representative crystal boules grown from the B-rich molten zone; (e) an X-ray micro-computed tomography cross-section image from the 3D scan of the boule; (f) optical microscope image of a optically polished sample cut from the crystal boule shown in (c). (scales in crystal photos are in mm)*

Since the stoichiometric composition doesn't allow crystal growth by melting at ambient pressure, the growth strategy was changed to one using a traveling solvent of non-stoichiometric composition to lower the melting point to control vaporization. As shown in Fig. 2(a), a Sc-rich flux melt of 55-65 at% Sc composition, and a B-rich flux melt of 82 at% composition should both stabilize precipitation of stoichiometric $ScB_2$ at temperatures below 2000 °C.

Multiple growths were performed from both the flux conditions, and the photographs of representative crystal boules are shown in Fig. 2(b-d). Microcomputed X-ray tomography ($\mu$-CT), reveal a significant inclusion of Sc-metal. Fig. 2(e) shows a 3D microcomputed cross-section of one of the SBS crystal boule (Fig. S3). Upon a careful observation one can clearly see the darker and brighter regions in the cross-section micrograph – the darker regions represent $ScB_2$ crystallites surrounded by brighter regions representing Sc metal-flux. Fig. 2(f) shows a photograph of cut and optically polished sliced sample from a SBB crystal boule (shown in Fig. 2(c)), at a 4 cm scale mark. This part of the crystal boule is a single grain and free from any visible inclusions. However, the outer surface of SBB crystal shown in Fig. 2(c) has a crust which is rich in boron (Fig. S4), the formation of such a crust resulted in the loss of B from the zone leading to the molten composition shifting towards stoichiometric composition leading to unstable growth condition and ultimately toward growth termination. The B-rich phase is tend to segregate out of the stoichiometric crystallites, Mayrhofer, et al., observed such behavior in $TiB_2$ thin films, the preferred oriented stoichiometric grains were found encapsulated in layers with excess in B.[30] The thinner regions of crystal boule are the resultant of such growth instabilities. Overall, it resulted in a sizable inclusion free crystal boule, and its composition and crystal quality are analyzed below.

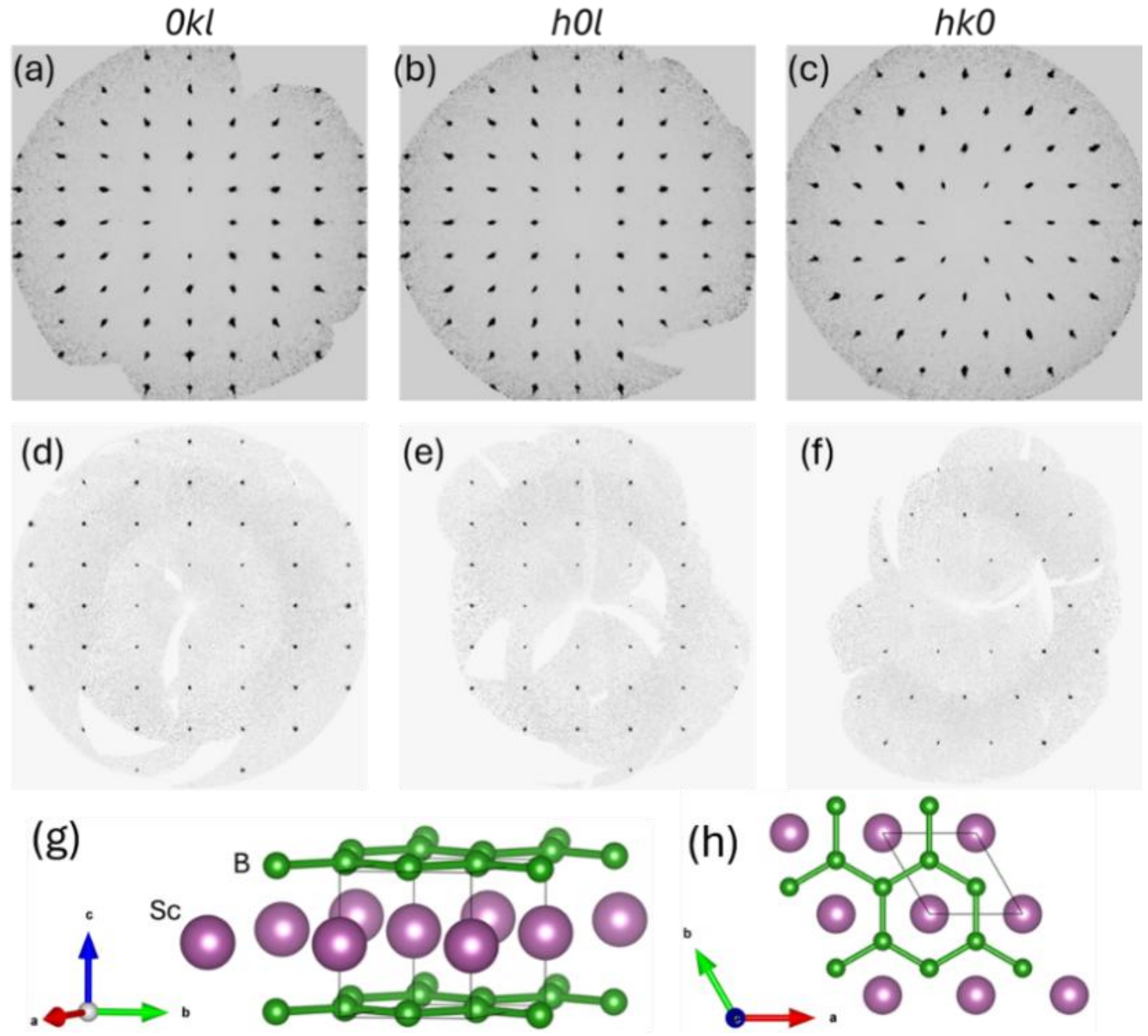

*Figure 3. Single crystal X-ray diffraction unwarp reciprocal space slices for (0kl), (h0l), and (hk0) planes for SBS (a-c) and SBB (d-f) crystals. The $ScB_2$ crystal lattice: (g) is the side view of Sc and B layers, and (h) is view from the top of layers along c-direction.*

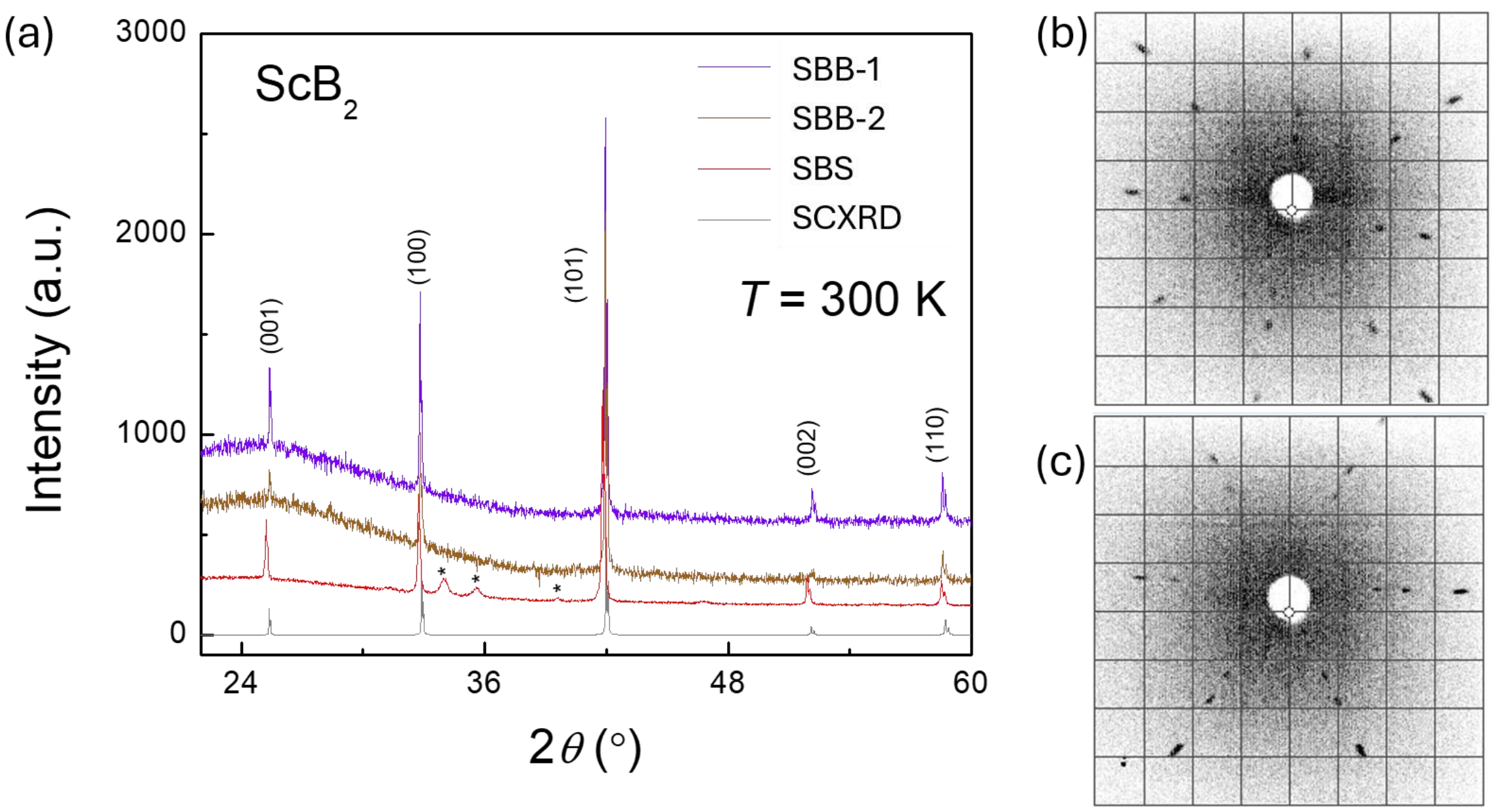

*Figure 4. (a) The powder X-ray patterns for two SBB crystals (purple & brown) and a SBS crystal (red), recorded at room temperature. Calculated powder diffraction pattern (black) from a CIF file for SCXRD structure solution of a sample from SBS. The small peaks in the red curve marked with asterisks belong to the Sc metal inclusions in Sc-flux grown crystal. (b) and (c) are the back-reflection Laue patterns for [100] and [001] directions on one of the B-rich molten grown crystals. The [100] are along the growth direction and [001] are perpendicular to the crystal length.*

## Crystal structure

The SCXRD unwarped slices for (0kl), (h0l), and (hk0) planes for the SBS and SBB crystals respectively in Fig. 3(a-c) and Fig. 3(d-f). All patterns clearly reflect a hexagonal lattice symmetry. The full SCXRD data set recorded at $T$ = 213 K for SBS crystal was analyzed and analysis parameters are given in **Table 1**. The structure refinement was fully converged, and the values of R parameters are very small in the best fitting range. The side and out of plane views of the solved crystal structure in Fig 3(g, h), show the close-packed triangular and honeycomb layers of Sc and B, respectively, as in an $AlB_2$-type structure. The evaluated lattice constants for SBS (resp. SBB) crystals are comparable $a$ = 3.1423(2) Å (resp. 3.1502(3) Å) and $c$ = 3.5084(3) Å (resp. 3.5041(3) Å), though the SBS crystal has a larger c/a ratio than the SBB one, likely indicative of a small compositional driven change in defect chemistry. The lattice constants for the SBB crystal are much closer to the previously reported room temperature values, $a$ = 0.314820(3) Å, $c$ = 0.351483(5) Å.[12,22,24,31]

The phase purity of both types of crystals was analyzed by powder x-ray diffraction, and the room temperature diffraction patterns are shown in Fig. 4(a). The black plot is calculated powder x-ray diffraction pattern by using a CIF file obtained of the solved crystal structure for SBS. The red plot of for the sample from SBS crystal and, evidently, in addition to the main $AlB_2$ phase, it has Sc metal peaks suggesting the inclusion of metal flux between $ScB_2$ grains, consistent with the $\mu$CT measurements. The purple and brown patterns are for the samples taken from the two different parts of a SBB crystal. All the peaks in these patterns are in good agreement with the calculated pattern. The SBB crystals have only single phase in the measured 2-theta range. Fig. 4(b,c) are the back-reflection Laue patterns along [100] and [001] directions for a SBB crystal. The [100] is along the growth direction and [001] is perpendicular growth direction.

Table 1. Crystal structure analysis parameters for SBS

| Chemical formula | $B_2Sc$ |
|---|---|
| $M_r$ | 66.58 |
| Crystal system, space group | Hexagonal, $P6/mmm$ |
| Temperature (K) | 213 |
| $a$, $c$ (Å) | 3.1423 (2), 3.5084 (3) |
| $V$ (Å$^3$) | 30.00 (1) |
| $Z$ | 1 |
| Radiation type, wavelength | Mo $K\alpha$, 0.71073 (Å) |
| $\mu$ (mm$^{-1}$) | 5.24 |
| $T_{min}$, $T_{max}$ | 0.641, 0.917 |
| No. of measured, independent and observed [$I > 2\sigma(I)$] reflections | 892, 38, 38 |
| $R_{int}$ | 0.026 |
| $(\sin \theta/\lambda)_{max}$ (Å$^{-1}$) | 0.749 |
| $R[F^2 > 2\sigma(F^2)]$, $wR(F^2)$, $S$ | 0.011, 0.025, 1.30 |
| $\Delta\rho_{max}$, $\Delta\rho_{min}$ (e Å$^{-3}$) | 0.24, -0.37 |

**High-resolution X-ray diffraction**

HRXRD measurements were performed on freshly cut and optically polished surfaces of SBB single crystals to assess lattice quality, mosaicity, defect structure. Fig. 5(a) shows the high-resolution $2\theta$-$\omega$ scan recorded on sample aligned for ($h$000) reflections. Only the expected (10-10), (20-20), and (30-30) peaks are observed, with no addition reflections or splitting, confirming that the crystals are single phases $ScB_2$.

Rocking curves for (10-10) and (20-20) planes (Fig. 5b-c) exhibit single well-defined peaks with FWHM of 255 arcsec and 132 arcsec (<0.1°), placing the crystal in free from grain boundaries and low-mosaic regime [31]. Although there is no direct literature on $ScB_2$ rocking curves for (h000) planes, these values are somewhat larger than those reported for the seeded growth $ZrB_2$ crystals [31],[18] suggesting the presence of a modest density of local defects or small in-plane angular deviations. In general, in transition metal diborides, the B vacancy defects with lowest formation energy most favorable over interstitial, anti-site, and even metal vacancy defects.[32] The narrower higher-order (20-20) reflection indicates excellent long-range periodicity. In both the RCs, the scattered intensity extends more prominently on the $+\omega$ side of the peak, over a relatively wide angular range. This is characteristic of local microstrain or point-defect-induced compressive lattice distortion rather than large-scale mosaic tilt. Under B-rich flux growth conditions it is plausible that the dominant point defect is Sc vacancies. Such defects would locally create stress in the lattice around the defect core, enhancing scattered intensity along the tails at higher angles, about RC peak. An alternative, less likely, explanation is that such defects are instead caused by B atom interstitials.[32–35] Regardless, HRXRD, however, cannot uniquely identify defect species in this case, and warrants future follow-up work through complementary microscopic characterization.[30,33,34]

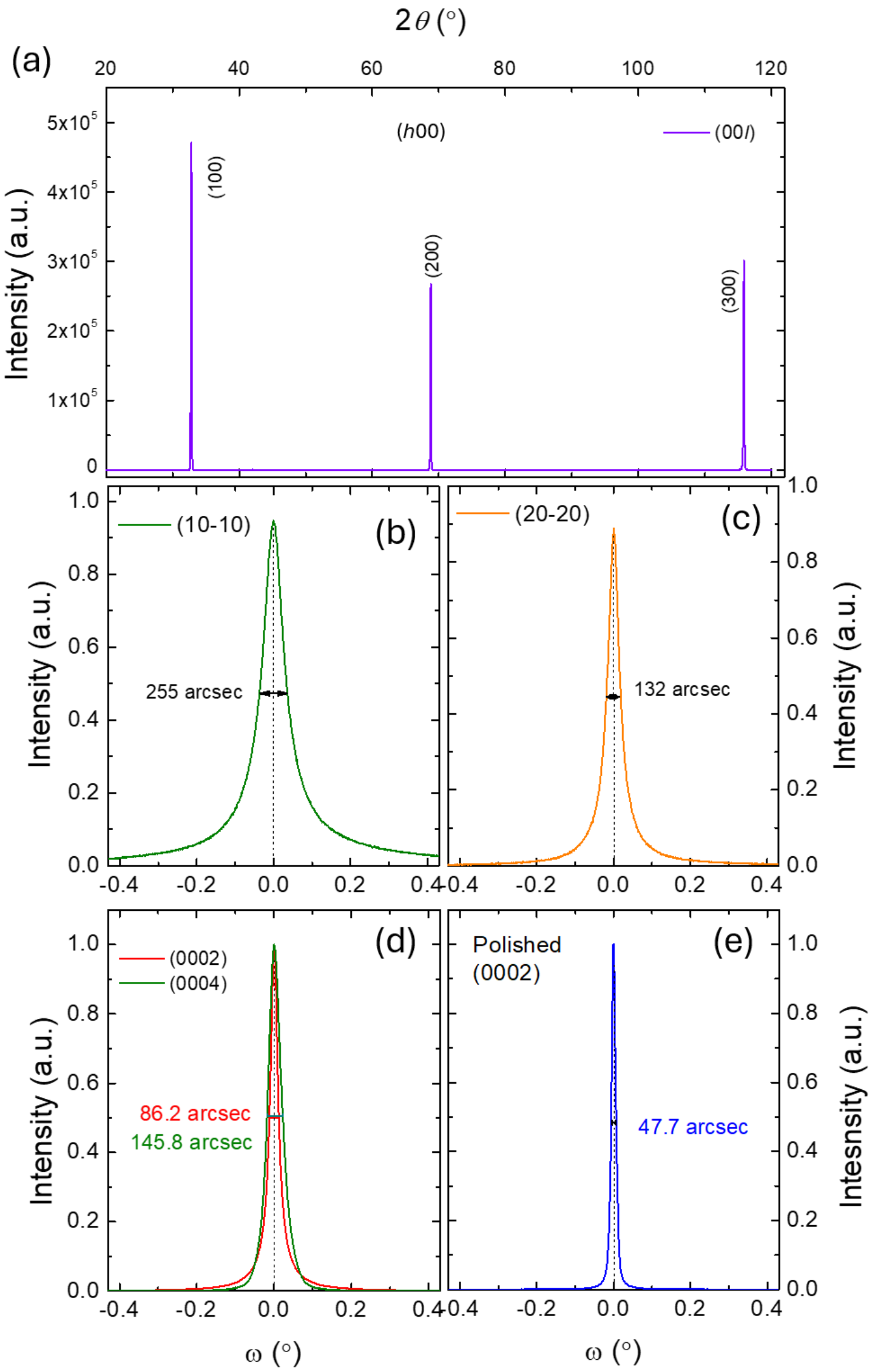

*Figure 5. (a) The high-resolution 2θ–ω x-ray diffraction pattern recorded on the freshly cut (h00) surface of a sliced circular SBB sample. (b, c) The ω rocking curves recorded for (10-10) and (20-20) diffraction planes on the optically polished surface. (d) The RCs for (0002) and (0004) diffraction planes recorded on unpolished surfaces, and (e) is the RC for (0002) diffraction planes on polished surface of SBB crystal.*

In order to assess the quality of basal planes of Sc and B layers along the c-axis, crystals were aligned and cut for (000*l*) planes. RCs for (0002) and (0004) reflection on freshly cut, unpolished surface are shown in Fig. 5(d). The FWHMs are 86.2 arcsec for (0002) and 145.8 arcsec for (0004), indicating high out-of-plane order and a well-defined stacking sequence of layers. The single-peak nature of these RCs again confirms that the crystal exhibits no grain boundaries or large-angle misorientation along the out-of-plane 000*l*) axis. The FWHM of the (0004) reflection on this unpolished crystal is comparable to that of the (20-20) reflection of polished crystal, reinforcing the conclusion that the bulk lattice is of high crystalline quality. Upon polishing, the RC for (0002) narrowed down significantly as shown in Fig. 5(e), and its FWHM of 47.7 arcsec showing that most tail intensity in unpolished samples originates from surface roughness rather than bulk disorder. The RC is highly symmetric about its peak position, with only a slight excess intensity on the +ω side, again consistent with a low density of compressive point defects rather than extended large-scale defects. This value of FWHM is comparable to seeded grown $ZrB_2$ crystals.[18] The SCXRD refined lattice constants of SBB closely match reported $ScB_2$ values,[22,23] further confirming the high crystalline quality. In thin films of some transition metal-diborides formation of antiphase boundaries have been studied, experimentally and theoretically.[34]

Taken together, the HRXRD results for both (*h*000) and (000*l*) planes show that SBB crystals possess highly coherent Sc and B layer stacking along c-axis and very low mosaicity, with only minor in-pane misorientation. These crystals are nearly perfect, with a modest density of local point defects, and this structural state provides an ordered hexagonal basal plane suitable for lattice-matched epitaxy AlGaN.

**Composition and Microstructure Analysis**

Microstructural and compositional analysis on SBS and SBB crystals were performed through EDS and EBSD. Prior to mechanical or ion polishing, fresh cleaved surfaces were analyzed with EDS (Fig. S5). After final surface polishing with the ion beam, regions from SBS and SBB were mapped with EDS and EBSD. The SBB sample shows a smooth surface uniform contrast in SEM imaging, with a uniform spatial distribution of Sc and B in EDS map from this region as shown in Fig. 6(a). Weak Ar peaks from Ar BIB polishing are present in the spectrum in Fig. S6. Conversely, Fig. 6(b) shows that the SBS sample contains a mix of regions containing Sc and B, and regions containing Sc metal, O, and C. Oxides likely formed in the excess Sc metal regions, whereas the source of C is likely from the use of a binder or from mechanical polishing.

Microstructural analysis of these regions mapped with EDS via EBSD confirms the single crystal nature of the $ScB_2$ of both the SBS and SBB samples, though they exhibit distinct primary crystallographic orientations of [10-10] and [-12-11], respectively (Fig. 7(a,b,e,f)). Band contrast (Fig. 7(c)) and Image Quality maps (Fig. 7(d)) provide insight into the pattern and surface quality of the mapped regions, where brighter grayscale values are generally higher surface and pattern quality, and darker grayscale values are from lower pattern quality and rougher surfaces. The SBS sample presents with significant regions of low image quality, which are attributed to surface roughness in non-$ScB_2$ regions containing Sc metal, O, and C phases.

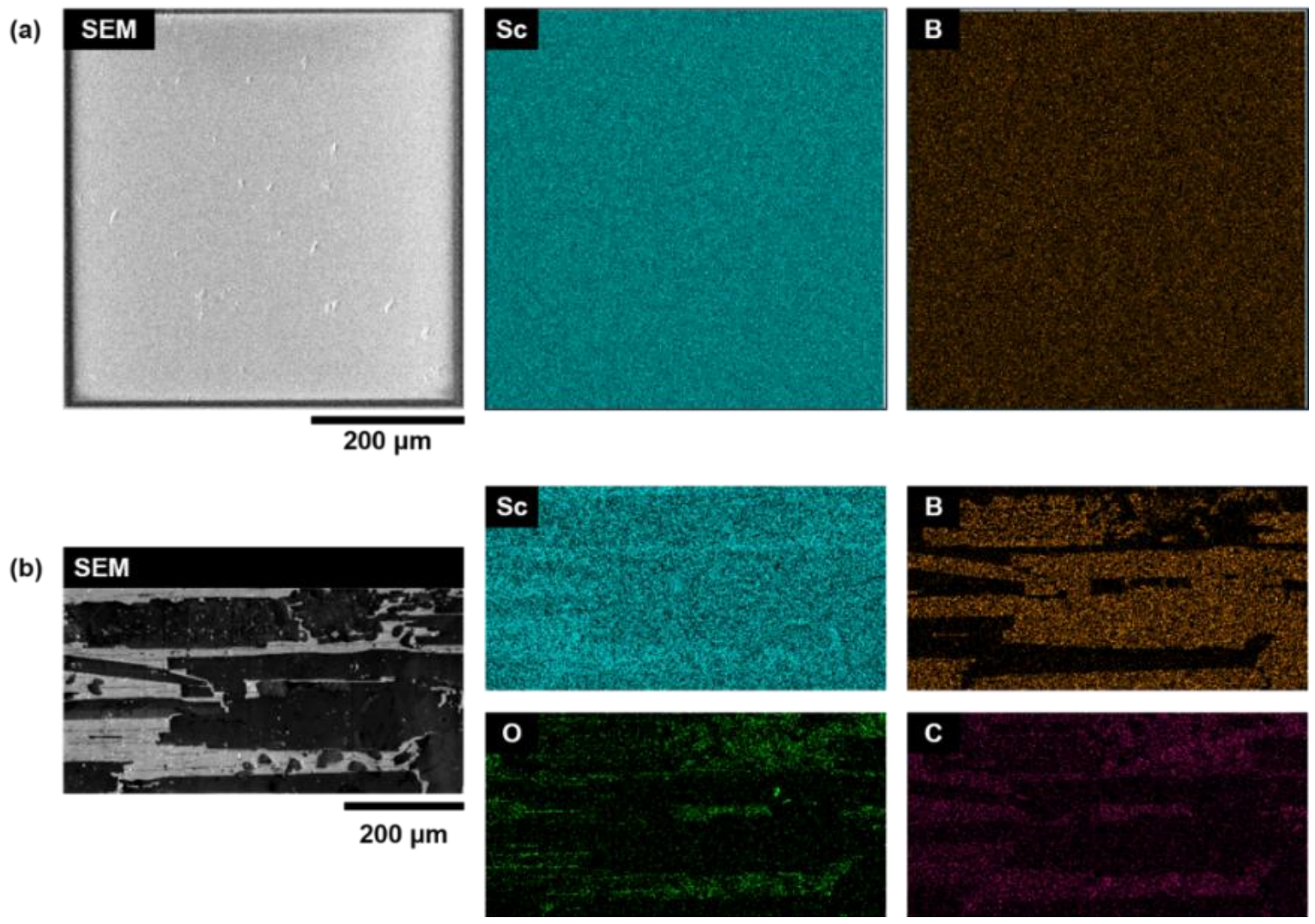

*Fig. 6. Secondary Electron SEM images and corresponding EDS maps of (a) SBB (b) SBS samples. The SBB crystal is composed of only $ScB_2$, whereas the SBS sample is composed of $ScB_2$ regions in a matrix of Sc metal, oxides, and carbides.*

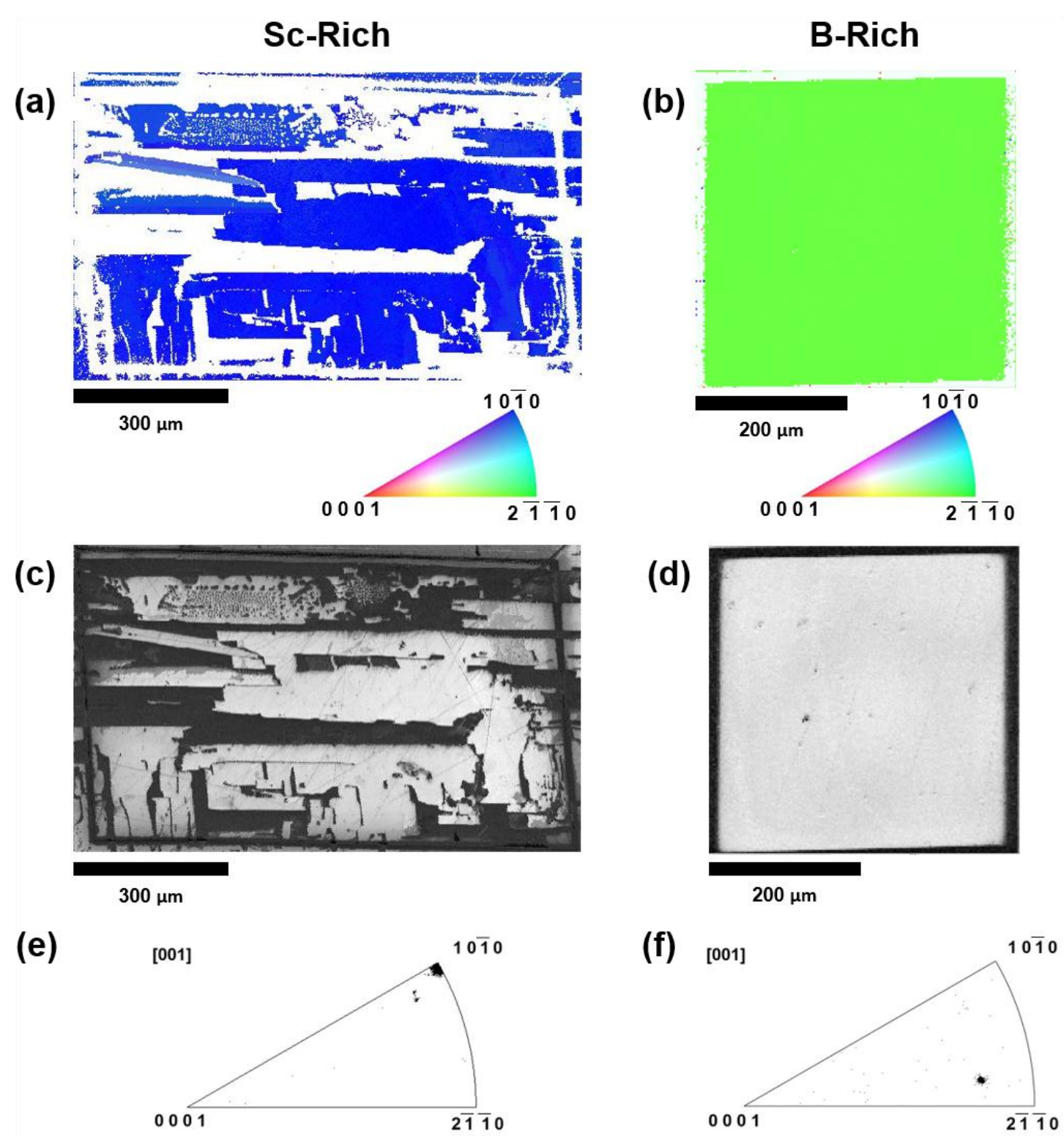

*Fig. 7. EBSD microstructural characterization of PFIB-milled rectangular boundary regions of SBS and SBB samples. (a) and (b) inverse pole figure (IPF) maps projected along the normal Z-direction show the primary crystal orientation is [10-10] for SBS and [-1 2 -1 1] for SBB. Maps of Image Quality in the SBS sample (c) and Band Contrast in the SBB sample (d) indicate good pattern quality in $ScB_2$ regions. The PFIB milled boundaries in each sample also show low quality (darker) and poor indexing. IPF pole figures are shown in (e) and (f), where orientations are tightly clustered towards near [10-10] and [-1 2 -1 1], respectively, consistent with single crystal structures.*

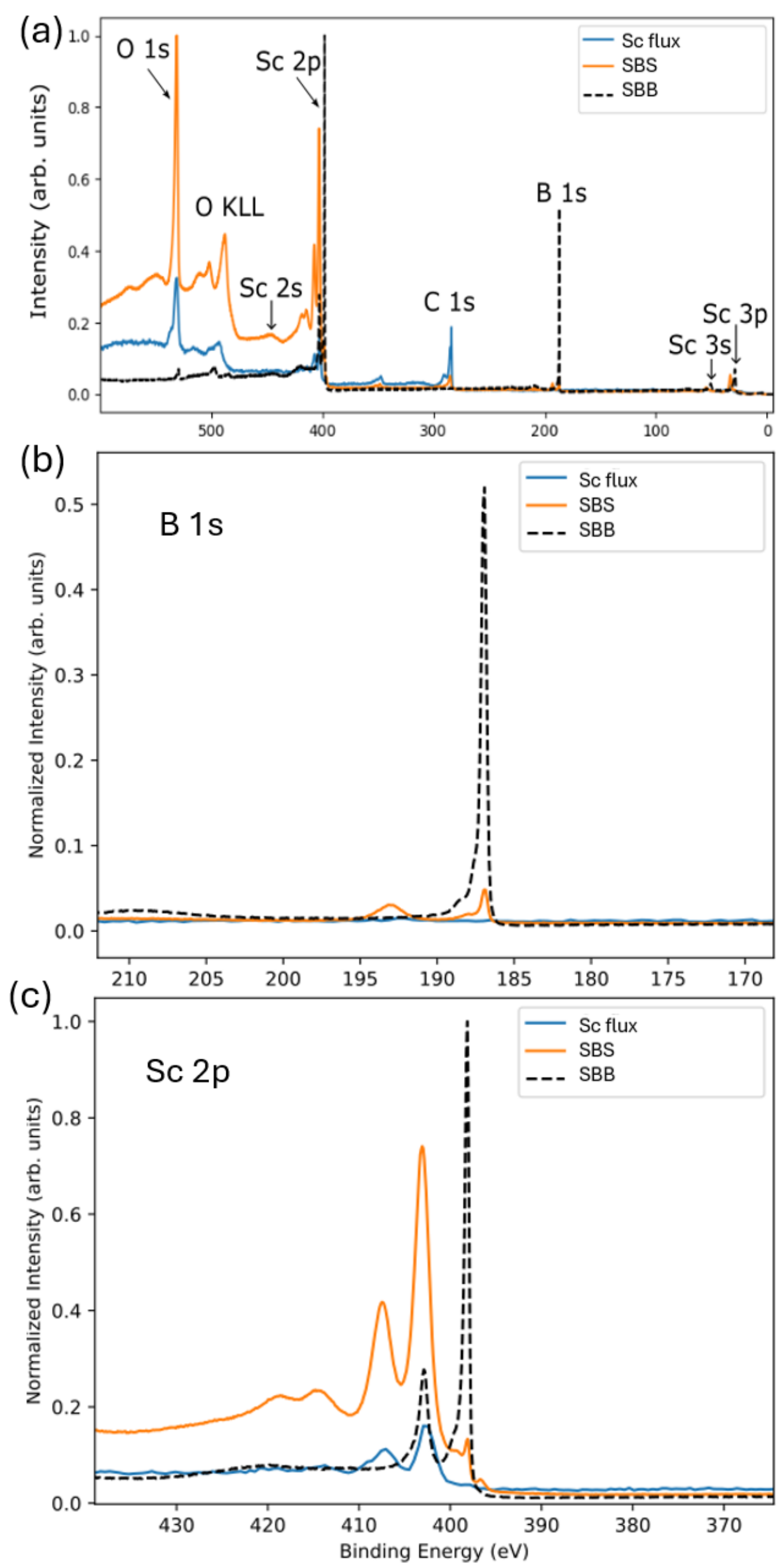

*Fig. 8. (a) XPS survey spectrum of both the SBS and SBB crystals with the binding energy peaks marked for B, Sc, O, and C. (b) and (c) represent the high-resolution elemental energy spectra of B 1s and Sc 2p for ScB2.*

## Electronic properties

The composition and electronic structure were measured using XPS for both the SBS and SBB as shown in Fig. 8. Using the 50 $\mu$m spot size we were able to map both $ScB_2$ crystallites and the Sc flux in the SBS sample and to confirm the uniformity of the single crystal SBB sample. The SBB single crystal was cleaved in UHV to reveal a fresh surface we can see from the overview spectra Fig. 8(a) that the only constituent elements are Sc and B. Because of the mix of $ScB_2$ and flux in the SBS sample, this sample did not cleave well, therefore we scratched the surface with a razor blade in situ to reveal a fresh surface. However, we still observe oxygen and carbon which present in both the Sc flux and $ScB_2$ crystallite regions. From the high-resolution B 1s we can see two distinct peak regions corresponding to $sp^2$ boron in the $ScB_2$ (187 eV) shoulders at 187.7 and 188.5 eV and $B_2O_3$ (193 eV).[36,37] The Sc 2p XPS compromised of the spin-orbit split double correspond to the $2p_{3/2}$ and $2p_{1/2}$.[38] For the regions dominated by the Sc flux, these are observed at 403.1 and 407.5 eV respectively in agreement with the formation $Sc_2O_3$ as a result of oxidation. The single crystal $ScB_2$ shows the same doublet shifted to lower binding energy (398.1 and 402.9 eV) corresponding to metallic Sc. The x-ray spot was larger than the crystallite and therefore the spectrum is comprised of the superposition of these two doublets in addition to a broader plasmon satellite at larger binding energies (414.3 eV and 418.7 eV). Work function measurements for the (110) plane is ~5 eV, suggesting the SBB crystal to be a good metal with extremely low electrical resistivity (~15 $\mu\Omega$-cm).[27]

**Summary**

In summary, we report the ambient pressure growth of $ScB_2$ single crystals using a self-flux traveling solvent floating zone method with both Sc-rich and B-rich fluxes. While Sc-rich flux conditions yield high-quality crystals, they contain significant Sc-flux inclusions. In contrast, B-rich flux conditions enable the growth of inclusion-free crystal boules with large dimensions of 4–5 mm in diameter and 30–50 mm in length. The B-rich flux approach resolves three major challenges in $ScB_2$ crystal growth: (i) the high vapor pressure of Sc metal, (ii) the formation of Sc-flux inclusions, and (iii) the difficulty of producing large, homogeneous crystals. Crystals grown by spontaneous nucleation from polycrystalline seeds under both flux conditions tend to grow along the [h00] direction within ±5–10°, exhibiting identical structures and compositions. Laue diffraction and HRXRD measurements show that SBB crystals grown under B-rich conditions form single blocks free of structural grain boundaries. Comparison of HRXRD rocking curves for the (h000) and (000l) planes indicates that the basal planes are of high quality, with FWHM values comparable to those reported for seeded $ZrB_2$ crystals. EBSD further confirms the highly aligned nature of $ScB_2$ crystal domains in SBS samples, while SBB crystals are single-domain. XPS analysis of the electronic structure reveals a high-quality lattice in SBB crystals, with a measured work function of ~5 eV, consistent with metallic behavior. The clean B 1s and Sc 2p spectra confirm that the Sc and B layers in SBB crystals are highly intact, providing a uniform hexagonal electrostatic potential at the surface suitable for AlGaN nucleation. Taken together, the growth, structural, and electronic characterization results establish $ScB_2$ as an ideal substrate candidate for 50% Al AlGaN UWBG semiconductors. The high structural quality of the (000l) planes offers opportunities for fabricating high-quality heterostructure interfaces with epitaxial AlGaN for vertical power devices. The high thermal and electrical conductivities of $ScB_2$ reported in our related work,[27] may further enhance the Baliga figure of merit by reducing substrate thermal and electrical resistances.[10] More broadly, the B-rich flux growth method shows promise for improving the crystal quality of other boride quantum materials that suffer from stoichiometry-related issues and sample inconsistencies—such as rare-earth borides ($SmB_6$, $YbB_{12}$, etc.). These materials are known for reproducibility challenges across the condensed-matter community,[39–41] and high-quality crystals may help advance understanding of their electronic and magnetic ground states, cooperative phenomena, and phase transitions.

**Acknowledgements**

This work was primarily supported as a part of APEX: A Center for Power Electronics Materials and Manufacturing Exploration, an Energy Frontier Research Center funded by the U.S. Department of Energy, Office of Science (growth, characterization, calculations analysis). Facilities management at Johns Hopkins University was supported by the National Science Foundation (Platform for the Accelerated Realization, Analysis, and Discovery of Interface Materials (PARADIM)) under Cooperative Agreement No. DMR-2039380. The Morgan State University authors acknowledge the Morgan Center for Education and Research in Microelectronics (MERC), supported by the State of Maryland, for equipment, space, and technical support staff. This work was authored in part by the National Laboratory for the Rockies (NLR) for the U.S. Department of Energy (DOE), operated under Contract No. DE-AC36-08GO28308. This research used high-performance computing resources at NLR. The XPS measurements were performed on APS beam time award(s) (DOI: https://doi.org/10.46936/APS-191174/60014913) from the Advanced Photon Source, a U.S. Department of Energy Office of Science user facility at Argonne National Laboratory. The views expressed in the article do not necessarily represent the views of the DOE or the U.S. Government.

# Supplementary Information

## Table of Contents

**Melt test of stoichiometric polycrystalline $ScB_2$**

(a)

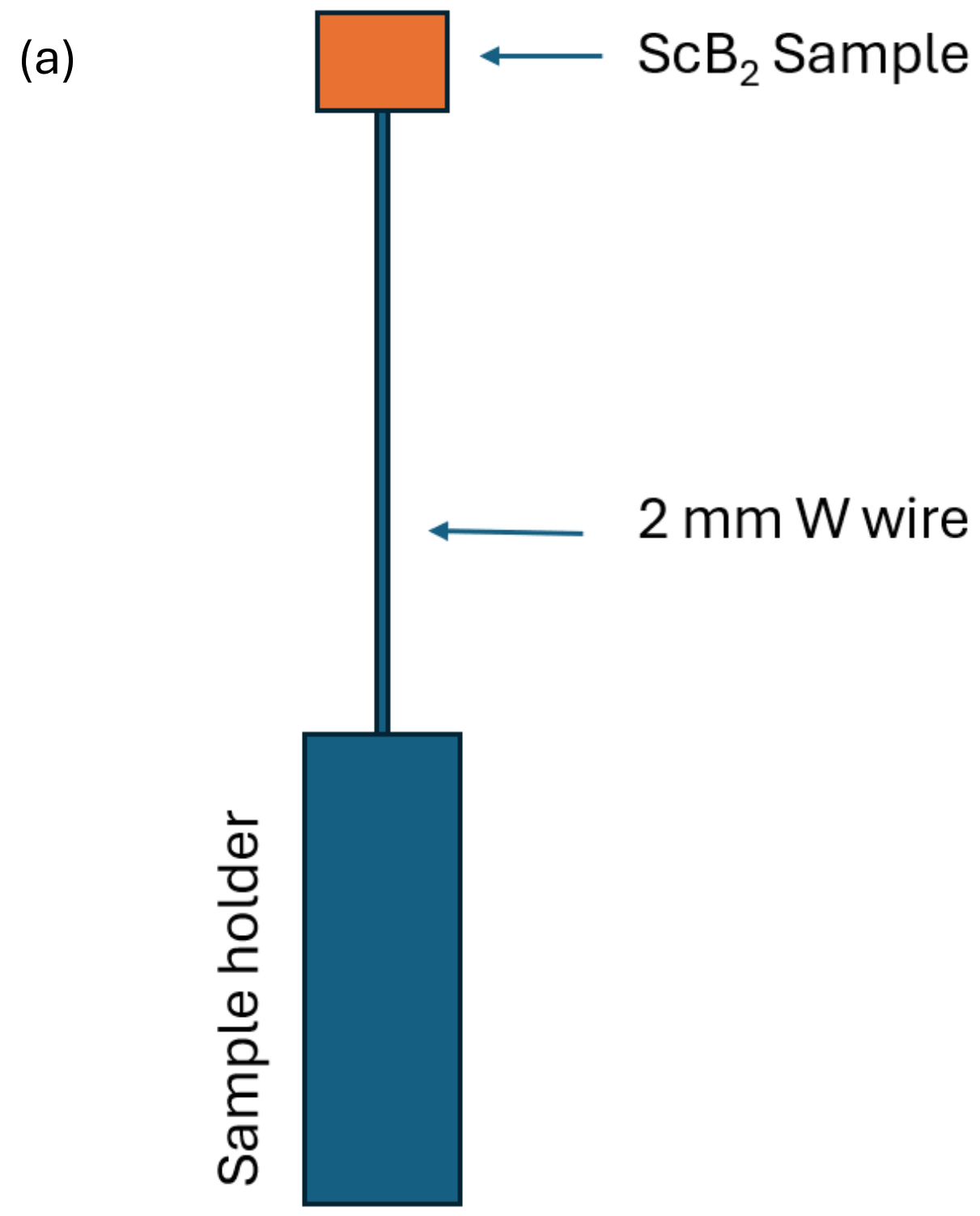

(b)

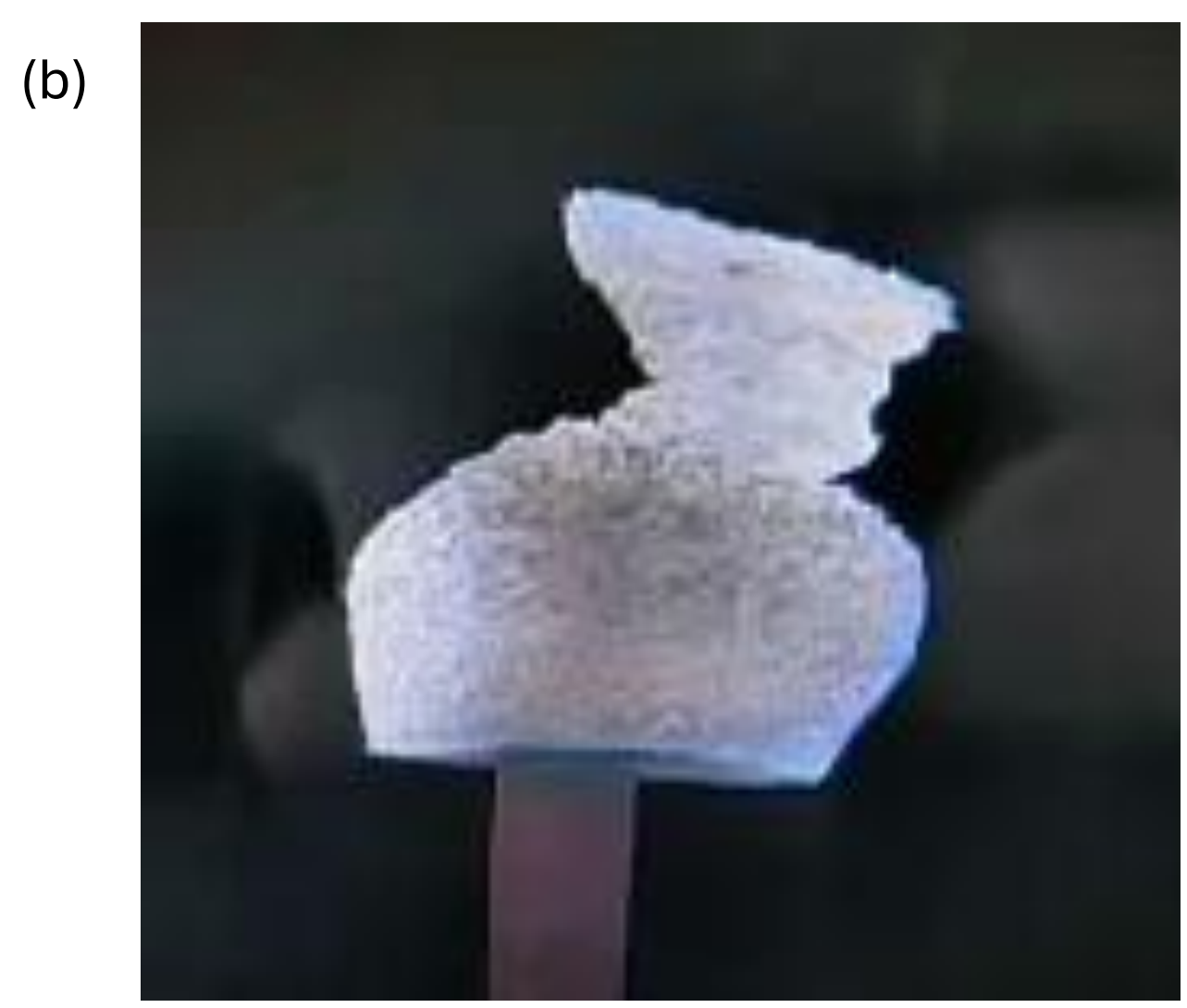

***Fig. S1.*** *(a) Schematic for the melt test of a small ScB2 piece, held on a 2 mm W wire. This thin diameter of W wire restricts the transport of heat away from the ScB2 sample and allows heating to high temperatures. The W wire is held on stainless steel sample holder. (b) The photo of sample after melt-test at ~98% laser power heating.*

**SEM analysis of melt-test sample**

To study the freshly cleaved surfaces, a Scanning Electron Microscope measurements were done on ZEISS EVO10 system with an accelerating energy of 15 keV in the Secondary Electron (SE) mode. Energy Dispersive X-ray analysis and elemental mapping were performed on SmartEDX microanalysis system which is optimized for the detection of lighter elements like Boron, Carbon and Oxygen. Back Scattered Electron (BSE) image was taken at a working distance maintained between 7.5 – 8 mm.

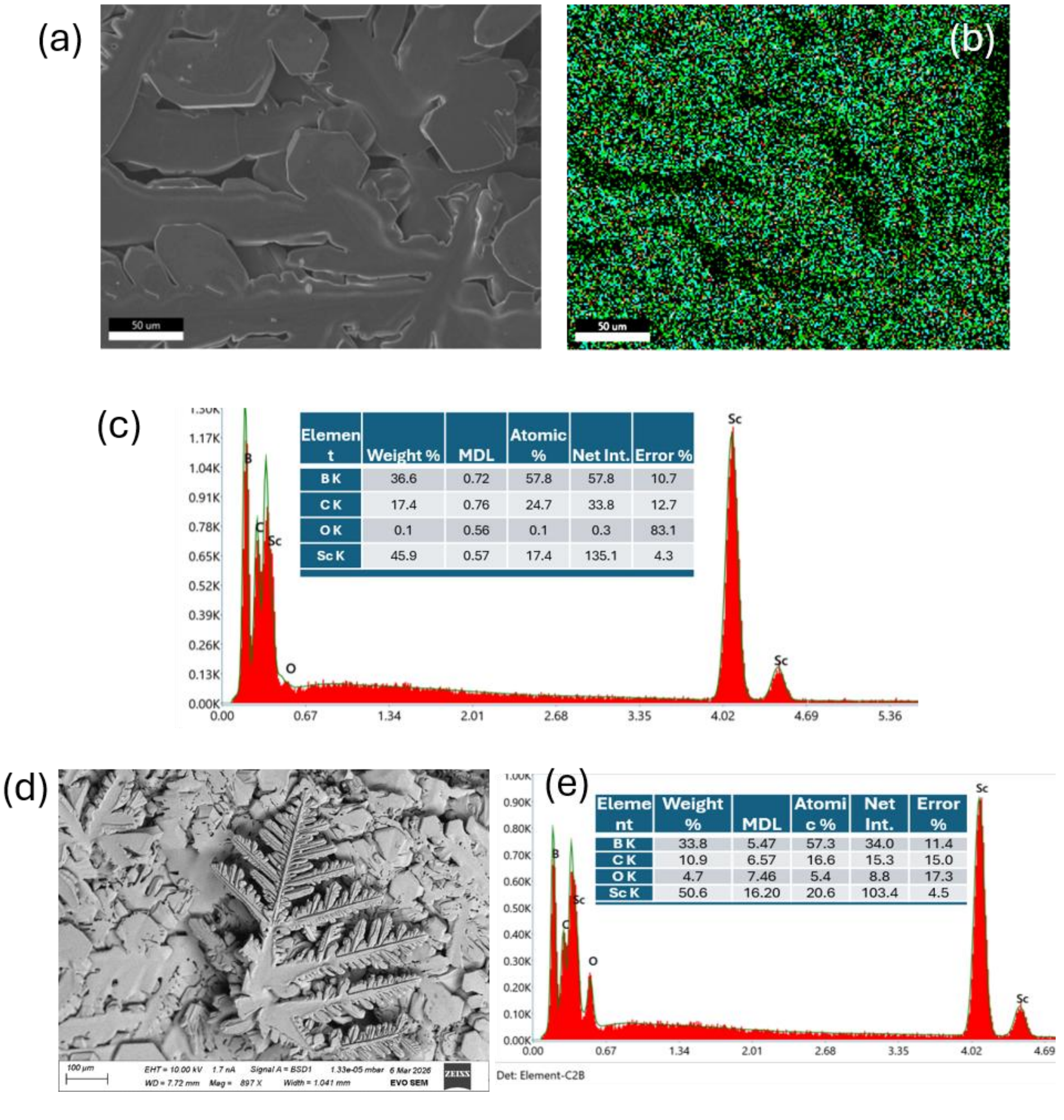

| Element | Weight % | MDL | Atomic % | Net Int. | Error % |
|---|---|---|---|---|---|
| B K | 36.6 | 0.72 | 57.8 | 57.8 | 10.7 |
| C K | 17.4 | 0.76 | 24.7 | 33.8 | 12.7 |
| O K | 0.1 | 0.56 | 0.1 | 0.3 | 83.1 |
| Sc K | 45.9 | 0.57 | 17.4 | 135.1 | 4.3 |

| Element | Weight % | MDL | Atomic % | Net Int. | Error % |
|---|---|---|---|---|---|
| B K | 33.8 | 5.47 | 57.3 | 34.0 | 11.4 |
| C K | 10.9 | 6.57 | 16.6 | 15.3 | 15.0 |
| O K | 4.7 | 7.46 | 5.4 | 8.8 | 17.3 |
| Sc K | 50.6 | 16.20 | 20.6 | 103.4 | 4.5 |

***Fig. S2.*** *SEM+EDS analysis of the sample after a melt test: (a) SEM image of the top surface melt-test sample showing the deposited material that is sublimated during heating. (b) and (c) the EDS areal map and pattern for the elemental distribution, a table insert in (c) shows the elemental ratio. (d) shows the SEM image of dendrite flakes on the top of the melt-test sample, the EDS spectra in (e) and a table insert indicates that these dendrites are B-rich ScB2.*

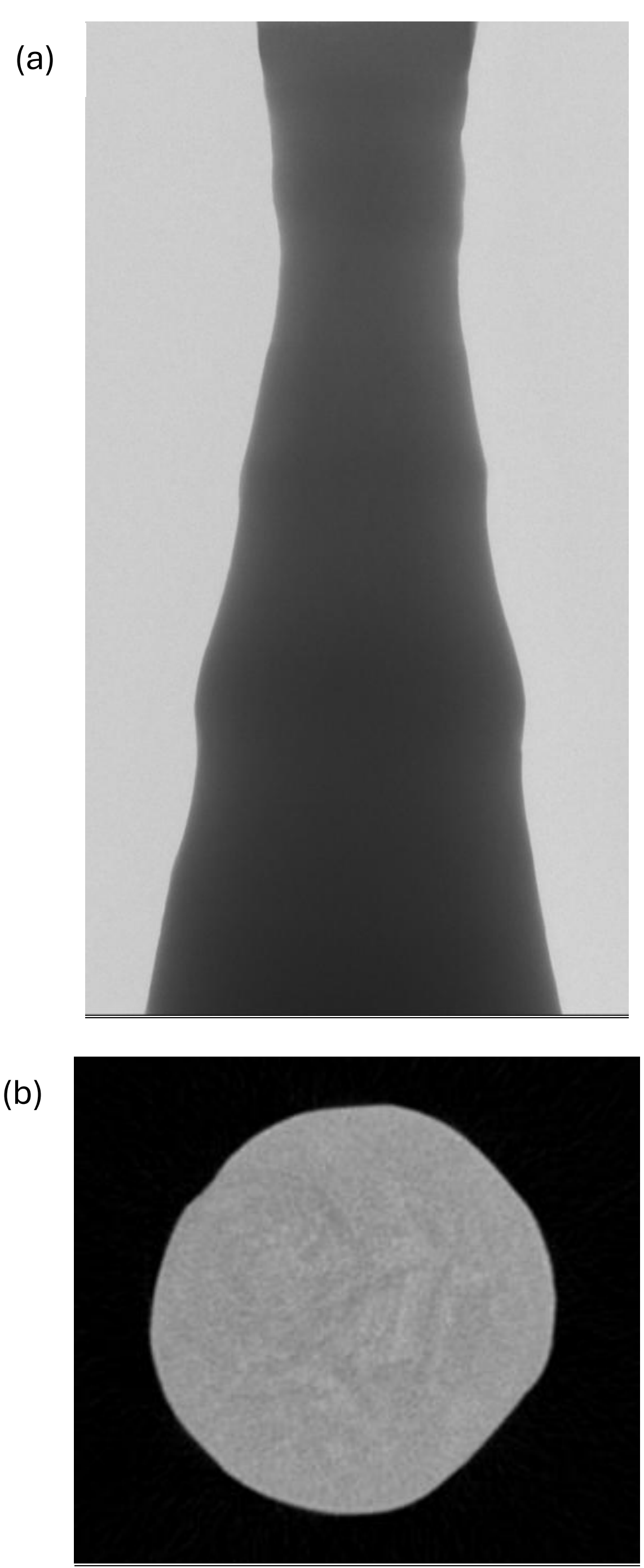

***Fig. S3.*** *(a) side and (b) cross-sectional views of a microcomputed X-ray tomography (μ-CT) 3D maps recorded for the SBS crystal boule.*

## SEM & EDS analysis of freshly cleaved samples

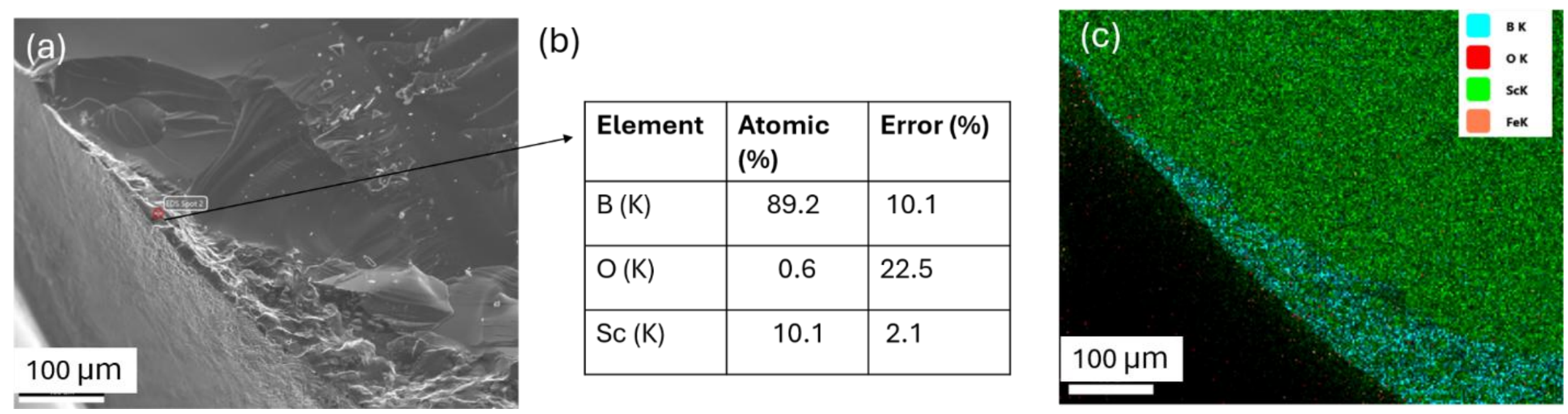

| Element | Atomic (%) | Error (%) |
|---|---|---|
| B (K) | 89.2 | 10.1 |
| O (K) | 0.6 | 22.5 |
| Sc (K) | 10.1 | 2.1 |

***Fig. S4.*** *(a) SEM and EDS analysis of one of the freshly cleaved SBB crystal, exhibiting a B rich crest on the outer layer of crystal, (b) a table shows an elemental composition measured by a point measurement on the outer crust region. (c) the B and Sc distribution over the cleaved surface, exhibiting the outer layer is B-rich.*

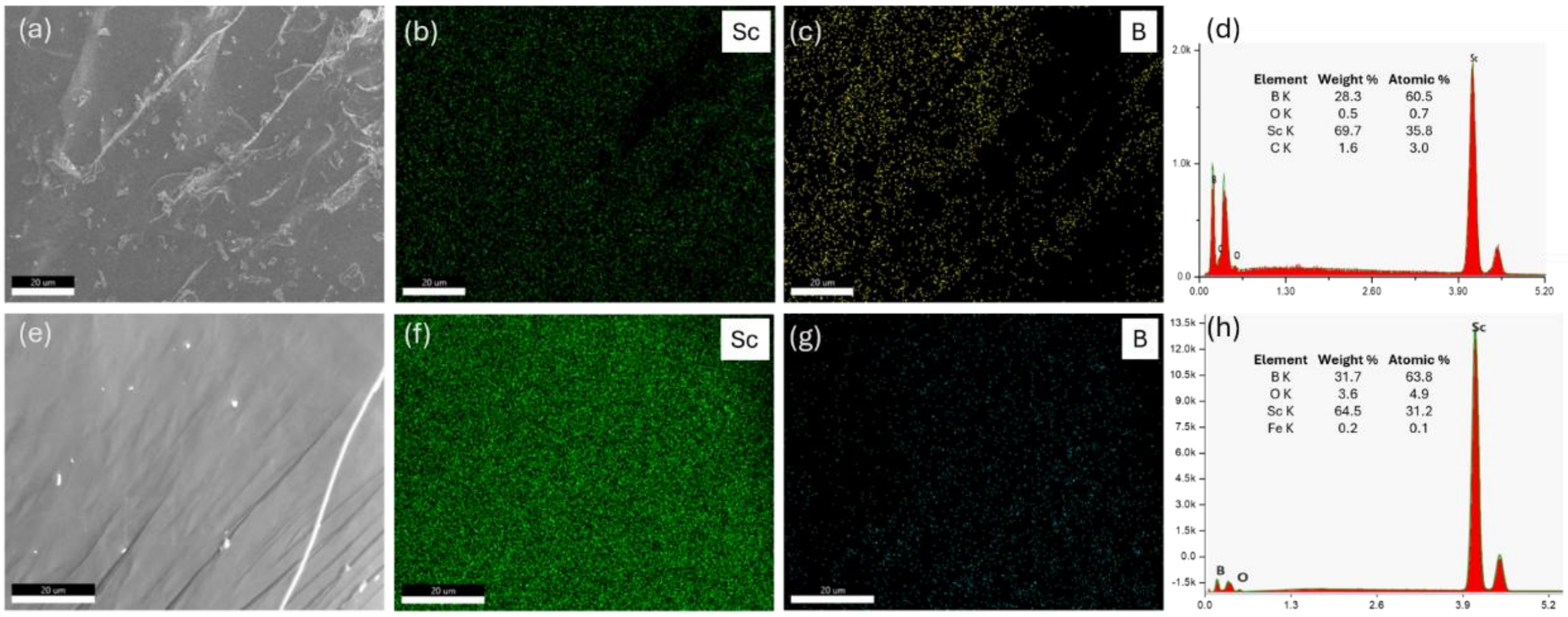

***Fig. S5.*** *A comparative elemental analysis of the freshly cleaved regions surfaces of ScB2 crystals, (a)-(d) SBS crystal grown from Sc molten, and (e)-(h) SBB crystal grown from B-rich molten.*

### EDS Spectra polished SBS and SBB samples

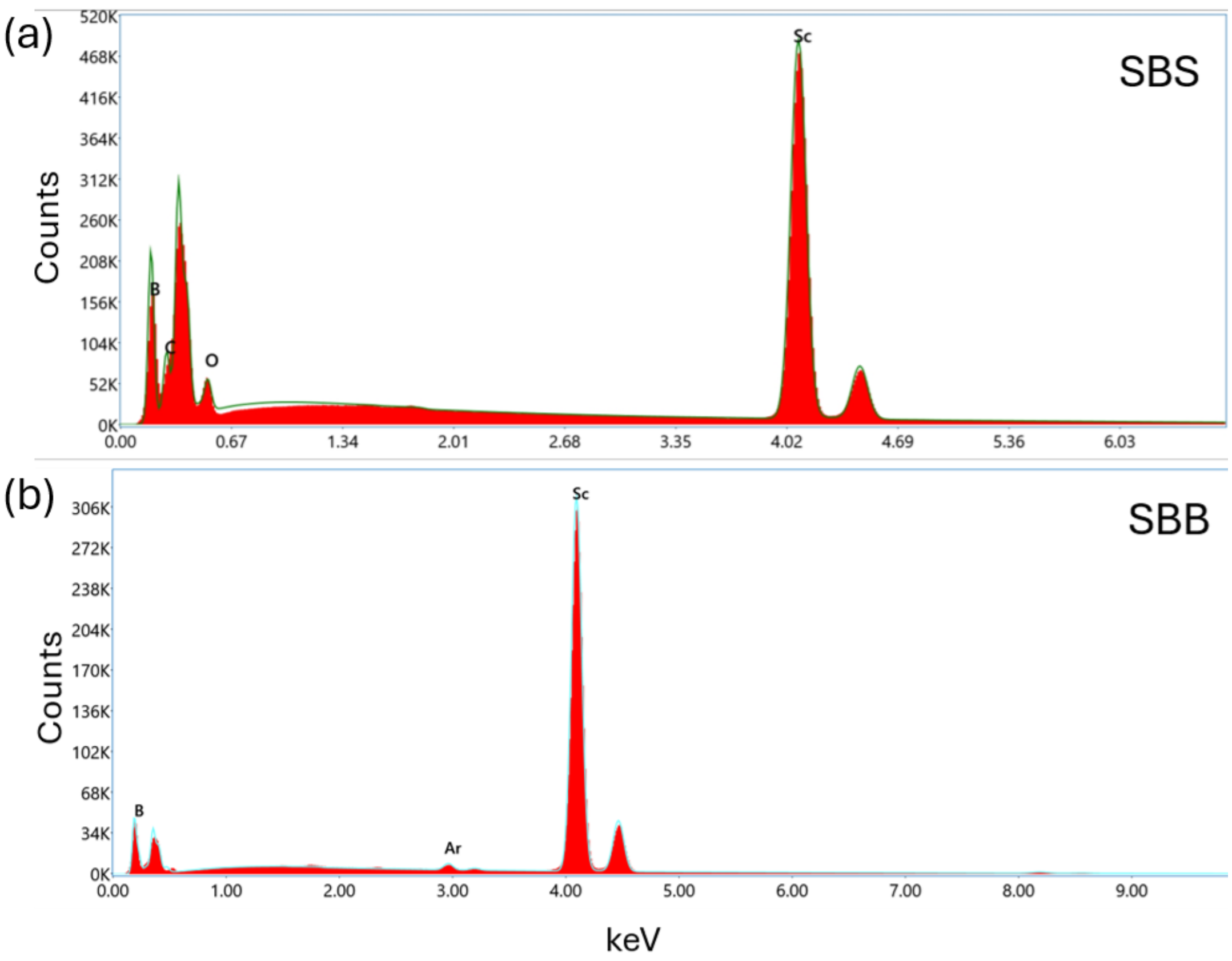

***Fig. S6.*** *The EDS patterns for the polished surfaces of SBS and SBB samples of the samples subjected to elemental and EBSD microstructure analysis in the main text and Fig. 6.*

## SEM Images of spherical voids

In some parts of one of the crystals, we observed upon optically polishing them. These spherical voids could result from the trapped gas at solid-liquid interface and ultimately became a part of the crystal lattice.

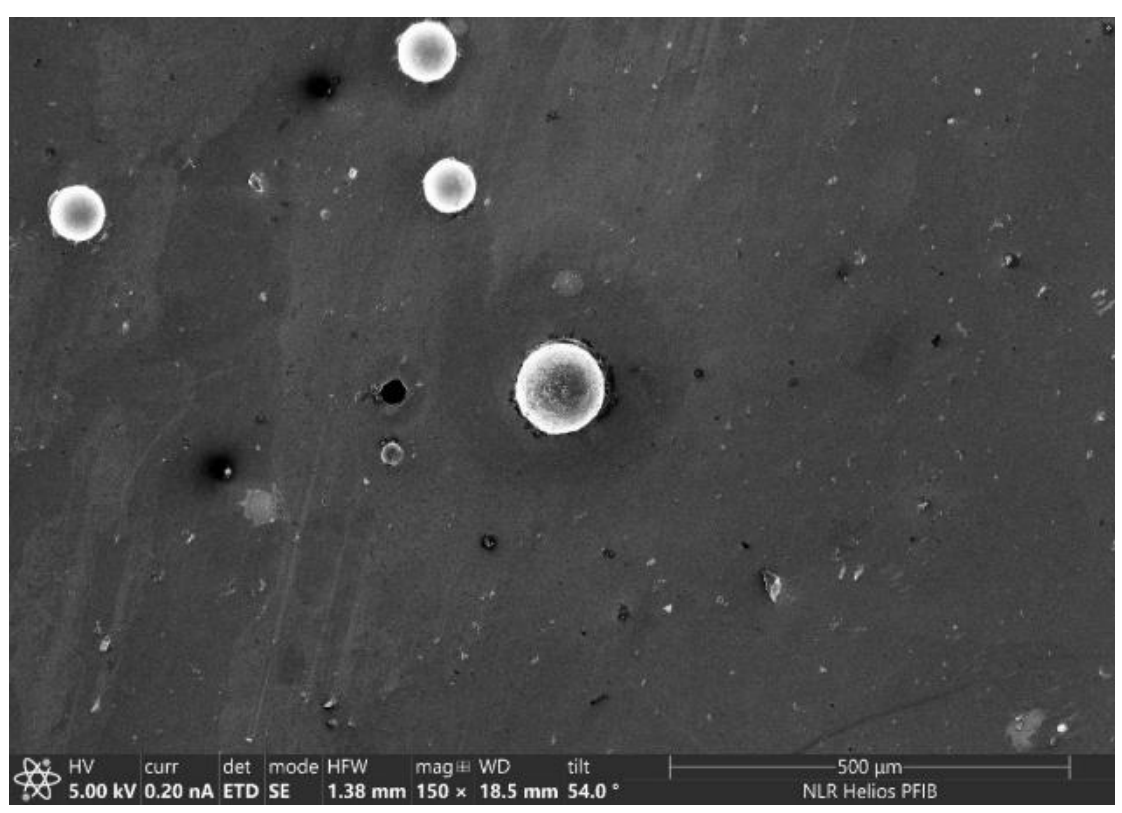

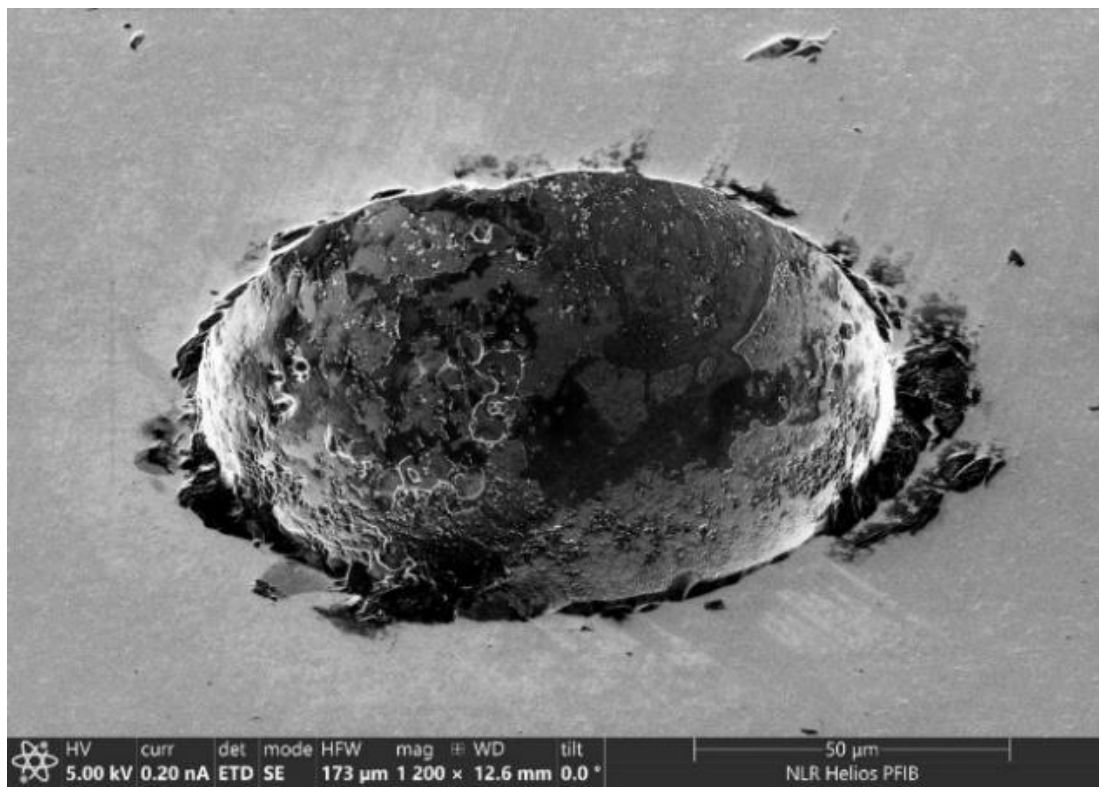

***Fig. S7.*** *SEM images showing the hemispherical voids in a particular region of a cut and optically polished SBB sample.*